\documentclass{article} 
\usepackage{iclr2026_conference,times}

\usepackage{amsmath,amsfonts,bm}

\def\eqref#1{equation~\ref{#1}}

\def\1{\bm{1}}

\DeclareMathAlphabet{\mathsfit}{\encodingdefault}{\sfdefault}{m}{sl}
\SetMathAlphabet{\mathsfit}{bold}{\encodingdefault}{\sfdefault}{bx}{n}

\usepackage{enumitem}
\usepackage{hyperref}
\usepackage{url}
\usepackage{booktabs}
\usepackage{multirow}
\usepackage{bm}
\usepackage{amsmath}
\usepackage{graphicx}
\usepackage{subcaption} 
\usepackage{wrapfig}
\usepackage{bbding}

\newcommand{\std}[1]{{\scriptsize $\pm$ #1}}

\title{RadDiff: Retrieval-Augmented Denoising Diffusion for Protein Inverse Folding}

\author{
Jin Han,\ Tianfan Fu,\ Wu-Jun~Li\thanks{Corresponding Author} \\
National Key Laboratory for Novel Software Technology\\ School of Computer Science, Nanjing University \\
\texttt{hanjin@smail.nju.edu.cn}, \texttt{\{futianfan,liwujun\}@nju.edu.cn}
}

\iclrpreprintcopy 

\begin{document}

\maketitle

\begin{abstract}
Protein inverse folding, the design of an amino acid sequence based on a target protein structure, is a fundamental problem of computational protein engineering. Existing methods either generate sequences without leveraging external knowledge or relying on protein language models~(PLMs). The former omits the knowledge stored in natural protein data, while the latter is parameter-inefficient and inflexible to adapt to ever-growing protein data.
To overcome the above drawbacks, in this paper we propose a novel method, called $\underline{\text{r}}$etrieval-$\underline{\text{a}}$ugmented $\underline{\text{d}}$enoising $\underline{\text{diff}}$usion~($\mbox{RadDiff}$), for protein inverse folding. In RadDiff, a novel retrieval-augmentation mechanism is designed to capture the up-to-date protein knowledge. 
We further design a knowledge-aware diffusion model that integrates this protein knowledge into the diffusion process via a lightweight module. 
Experimental results on the CATH, TS50, and PDB2022 datasets show that $\mbox{RadDiff}$ consistently outperforms existing methods, improving sequence recovery rate by up to 19\%.
Experimental results also demonstrate that RadDiff generates highly foldable sequences and scales effectively with database size.
\end{abstract}

\section{Introduction}
Proteins are essential biomolecules that carry out a wide range of biological functions within living organisms. In protein engineering, it is challenging to design novel proteins with desired functions. As the function of protein is determined by its three-dimensional~(3D) structure~\citep{koehler2023sequence}, protein inverse folding serves as a fundamental approach for functional protein design. The goal of protein inverse folding is to computationally design an amino acid sequence that will fold into a specified 3D protein structure~\citep{ingraham2019generative}.

Recent advances in deep learning have shown great promise for protein inverse folding~\citep{ingraham2019generative,jinglearning,hsu2022learning,fu2022sipf,dauparas2022robust,gaopifold,tan2023global}. Representative methods include diffusion-based methods~\citep{yi2023graph,bai2025mask}, which adopt the denoising diffusion model to generate amino acid sequences conditioned on the geometric features of protein structure. These models often operate \textit{de novo}, in which the generating process solely depends on the structure backbone~\citep{mahbub2025prism}. This process omits protein knowledge stored in natural protein data. Designing sequences without reference to known proteins may get sequences that are biologically suboptimal~\citep{huang2024interaction}.

Recognizing the value of protein knowledge, some methods have successfully improved protein design performance by incorporating information from pre-trained protein language models~(PLMs)~\citep{zheng2023structure, gaokw, wang2024diffusion}. PLMs can implicitly capture relevant amino-acid distributions after training on millions of natural protein sequences. While effective, these methods suffer from two key drawbacks. First, the PLMs often contain billions of parameters~\citep{hayes2025simulating}, resulting in parameter-inefficient architectures for downstream protein design tasks. Second, the knowledge in PLMs is static, which compresses the data into the fixed model parameters. As the protein data continues to expand rapidly, incorporating the up-to-date data requires retraining the entire PLM, which is both inflexible and computationally intensive. 

To address these challenges, in this paper we propose a novel method, called \underline{r}etrieval-\underline{a}ugmented \underline{d}enoising \underline{diff}usion~(\mbox{RadDiff}), for protein inverse folding. 
The main contributions of our work are outlined as follows:
\begin{itemize}
\item RadDiff designs a novel retrieval-augmentation mechanism to first retrieve a set of structurally similar proteins through hierarchical search, and then perform residue-wise alignment to identify matched structural regions. From the aligned residues, RadDiff constructs a position-specific amino acid profile that captures up-to-date protein knowledge.
\item RadDiff further designs a knowledge-aware diffusion model to incorporate protein knowledge from the amino acid profile through a lightweight module, making RadDiff more parameter-efficient than PLM-based methods. 
\item Experimental results on the CATH, TS50, and PDB2022 datasets show that RadDiff consistently outperforms existing methods, improving sequence recovery rate by up to 19\%. Experimental results also demonstrate that RadDiff generates highly foldable sequences and scales effectively with database size.
\end{itemize}

\section{Related Works} 
\label{sec:related} 
\subsection{Protein Inverse Folding}
Protein inverse folding has been extensively explored for years. Existing approaches can be grouped into two main categories: structure-only methods and knowledge-based methods.

\noindent\textbf{Structure-Only Methods.}
This category of methods uses only the protein structure as input, including both physics-based methods and deep learning–based methods. Physics-based  methods, such as Rosetta~\citep{alford2017rosetta}, formulate the problem as an energy minimization task. 
These methods are often computationally intensive, and their search space is limited.
More recently, deep learning-based methods, mainly including graph neural network~(GNN) based methods and diffusion-based methods, have shown great promise for this problem~\citep{wang2018computational,ingraham2019generative,jinglearning,qi2020densecpd,fu2022sipf,dauparas2022robust,gaopifold,yi2023graph,wang2024diffusion,qiu2024instructplm}. 
GNN-based methods, like GVP~\citep{jinglearning}, ProteinMPNN~\citep{dauparas2022robust}, and PiFold~\citep{gaopifold}, demonstrate the power of GNN to learn representations directly from protein structures. 
Diffusion-based methods, like GradeIf~\citep{yi2023graph} and MapDiff~\citep{bai2025mask}, have also shown great potential of using denoising diffusion models to generate the amino acid sequence. However, the above methods omit the protein knowledge stored in the natural protein data and design sequences with no prior knowledge.

\noindent\textbf{Knowledge-Based Methods.}
The knowledge-based methods not only take the protein structure as input, but also incorporate the protein knowledge from natural protein data. Methods such as LM-Design~\citep{zheng2023structure} and KW-Design~\citep{gaokw} successfully leverage pre-trained PLMs to inject protein knowledge into the design process, but these methods contain a large amount of parameters and are not aware of the up-to-date protein knowledge.
PRISM~\citep{mahbub2025prism} employs the concept of retrieval-augmentation. However, the core methods of PRISM are significantly different from RadDiff. PRISM operates at the embedding level, relying on pre-trained structure and sequence encoders to retrieve and integrate learned representations, which may lose the information of the original structure and sequence.

\subsection{Protein Structure Retrieval}
Protein structure retrieval aims to retrieve similar proteins from a protein structure database given a query structure. We introduce two classical protein structure retrieval methods, TM-align~\citep{zhang2005tm} and FoldSeek~\citep{van2022foldseek}, which will be used in our method.

TM-align~\citep{zhang2005tm} is a sequence-independent protein structure comparison tool. TM-align uses heuristic dynamic programming to find the optimal structure alignment between two structures based on template modeling score~(TM-score)~\citep{zhang2004scoring}. TM-score is a score function to measure the structure similarity between two protein structures, which has a value in (0,1], and 1 indicates that two structures are perfectly matched. TM-score$>$0.5 indicates that two structures are highly likely to share similar topology~\citep{xu2010significant}. We will use the version US-align~\citep{zhang2022us} in our method, which is an extension of the TM-align that can generate more accurate structural alignment. Although TM-align and US-align are accurate in identifying the similarity of two structures, they are time-consuming and infeasible to perform large-scale comparisons for millions or even billions of structure pairs. 

FoldSeek~\citep{van2022foldseek} is a fast protein structure retrieval method based on a structural alphabet. FoldSeek discretizes structures into 3D interaction (3Di) sequences and uses MMseqs2~\citep{steinegger2017mmseqs2} for ultra-fast retrieval, achieving search speed several orders of magnitude faster than traditional alignment-based methods like TM-align. However, due to the information loss in the discretization to a structural alphabet, FoldSeek is generally less accurate than TM-align~\citep{litfin2025ultra}. 

\section{Methods}
\begin{figure*}[t]
\centering
\includegraphics[scale=0.25]{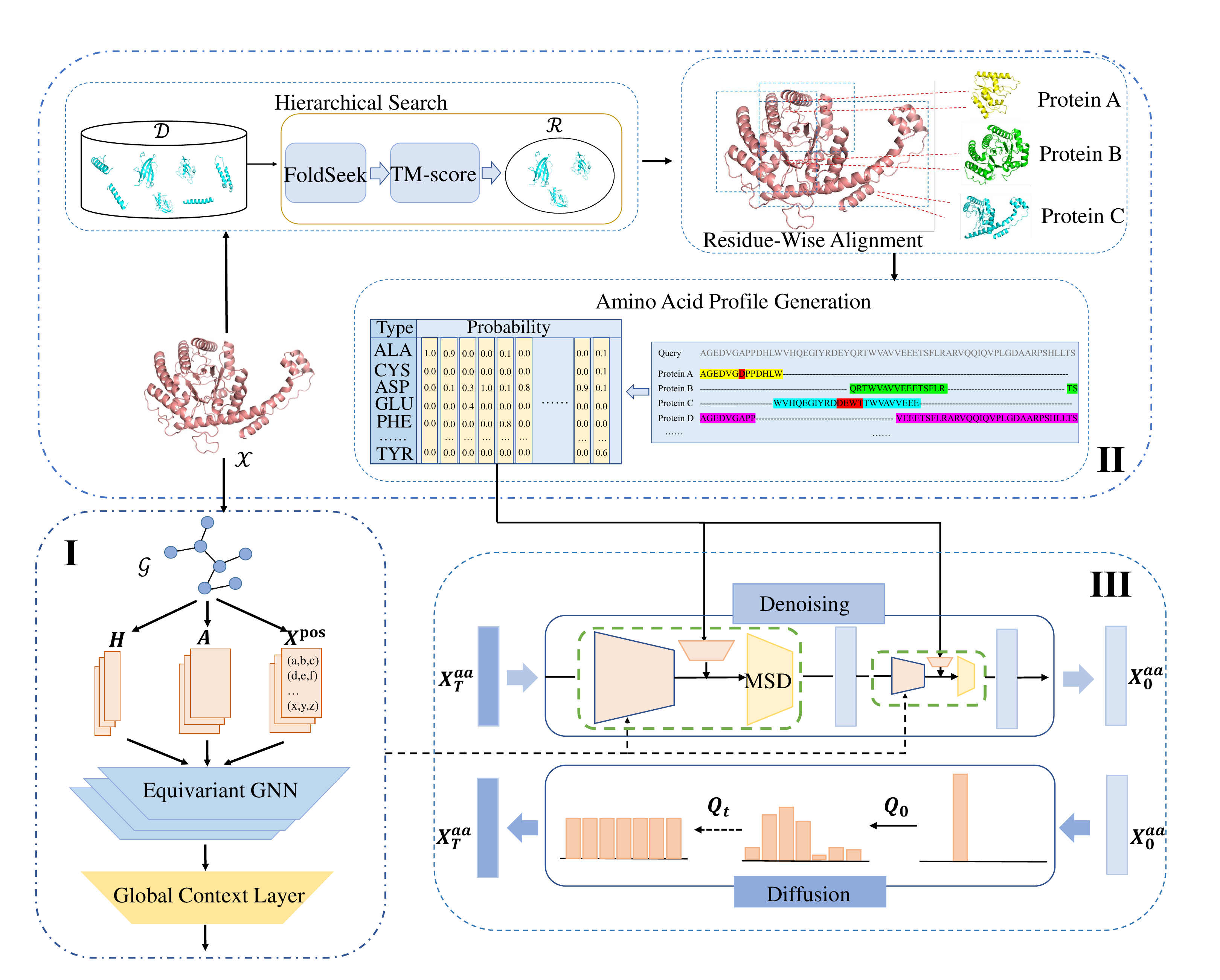}
\caption{RadDiff's architecture. (I) The input structure is encoded through an equivariant GNN and a global context layer to capture the geometric properties. (II) The same input structure is used to retrieve proteins with similar structures via a hierarchical search, followed by residue-wise alignment between the input structure and the retrieved set. This alignment generates an amino acid profile that reflects protein knowledge stored in natural protein data. (III) A knowledge-aware diffusion model is designed to integrate this protein knowledge into the diffusion process.}
\label{fig: methods}
\end{figure*}

RadDiff designs protein sequences not only based on the input protein structure, but also by leveraging a set of structurally similar proteins retrieved using the input structure. As shown in Figure~\ref{fig: methods}, the overall architecture of RadDiff consists of three components: a graph representation learning module that captures the geometric properties of protein structure, a novel retrieval-augmentation mechanism that captures protein knowledge from a set of retrieved proteins, and a knowledge-aware diffusion model that integrates retrieved protein knowledge into the diffusion process.

\subsection{Graph Representation Learning}
In RadDiff, we represent each protein as a residue-level graph and apply a graph neural network~(GNN) to capture both local and global structural features. 
\subsubsection{Graph Construction} 
We represent the protein structure as a graph $\mathcal{G=(V, E)}$, where each node $v_i\in \mathcal{V}$ corresponds to an amino acid. The graph's connectivity is defined using a k-nearest neighbor (kNN) algorithm constrained by a distance cutoff. In particular, an edge $e_{ij}\in \mathcal{E}$ exists between two nodes $v_i$ and $v_j$ only if their $C_\alpha$ distance is less than $30\mathring{\text{A}}$. The graph features consist of node features $\bm{H}$, coordinate features $\bm{X}^{pos}$, and edge features $\bm{A}$. The node features $\bm{H}$ contain the residue type, secondary structure, dihedral angles, solvent-accessible surface area~(SASA), crystallographic B-factor, and protein surface features~\citep{ganea2021independent, yi2023graph, bai2025mask}. The coordinate features $\bm{X}^{pos}$ are the node coordinates. The edge features $\bm{A}$ contain the relative spatial distance, local spatial positions, and relative sequential positions~\citep{bai2025mask}.

\subsubsection{Feature Updating}
\label{sec:3d_representation}
To represent 3D protein structure, we employ a global-aware equivariant graph neural network~(EGNN)~\citep{satorras2021n} as the network backbone.  The EGNN is composed of $L$ layers, where the $l$-th layer updates the node features $\bm{h}_i^l$ and coordinates features $\bm{x}_i^l$ while preserving SE(3) equivariance for each node $i$. Thanks to the SE(3) equivariance property, EGNN achieves equivariance to rotations and translations on the node coordinates. The $\bm{x}_i^0$ is $\bm{X}_i^{\text{pos}}$, and $\bm{h}_i^0$ is derived from the node feature $\bm{H}$.
At the $l$-th layer, the node and coordinates features are updated via a local message passing:
\begin{equation}
\begin{aligned}    
&\bm{m}_{ij}^l = \phi_e \left( \bm{h}_i^l, \bm{h}_j^l, \left\| \bm{x}_i^l - \bm{x}_j^l \right\|^2, \bm{a}_{ij} \right),\\
 &\bm{x}_i^{l+1} = \bm{x}_i^l + \frac{1}{|\mathcal{N}_i|} \sum_{j \in \mathcal{N}_i} \left( \bm{x}_i^l - \bm{x}_j^l \right) \, \phi_x \left( \bm{m}_{ij}^l \right), \\ 
&\bm{m}_i^l = \sum_{j \in \mathcal{N}_i} w_{ij}\bm{m}_{ij}^l,\\
&\bm{h}_i^{l+1} = \phi_h \left( \bm{h}_i^l, \bm{m}_i^l \right), \\ 
\end{aligned} 
\end{equation}
where $\mathcal{N}_i$ is the set of neighbors of node $i$ and $\phi_e, \phi_x, \phi_h$ are multi-layer perceptrons~(MLPs). $\bm{a}_{ij}$ is the edge feature between node $i$ and $j$. $\bm{w}_{ij}=\sigma(\phi_w(\bm{a}_{ij}))$, where $\sigma(\cdot)$ is the sigmoid function and $\phi_w$ is also a MLP.

We enhance this local message passing with a global context layer to allow for long-range communication across the protein structure~\citep{tan2023global, bai2025mask}. After the local update, the node representations are further refined as: 
\begin{equation}
\begin{aligned}
    &\bm{c}^{l+1} = \text{MeanPool}(\{\bm{h}_i^{l+1}\}_{i=0}^{N-1}), \\
  &\bm{h}_i^{l+1} = \bm{h}_i^{l+1}\odot\sigma(\phi_c(\bm{c}^{l+1}\Vert \bm{h}_i^{l+1})),
\end{aligned}
\end{equation}
where $\odot$ is the Hadamard product, $N$ is the number of nodes, $\Vert$ represents concatenation operation and $\phi_c$ is MLP. The above updating process is repeated for $L$ times. Finally, the output from the final layer, $\bm{h}_i^{L}$, is used as the representation of residue $i$.

\subsection{Retrieval Augmentation}
\label{sec:retrieval}
We use $\mathcal{P}$ to denote a protein, which contains its amino acid sequence $\mathcal{S}$ and 3D protein structure $\mathcal{X}$, i.e., $\mathcal{P}=(\mathcal{S}, \mathcal{X})$. The protein inverse folding problem aims to find a valid sequence $\mathcal{S}$ that folds into the desired structure $\mathcal{X}$. Unlike existing methods that only use protein structure as input, our method leverages the knowledge from existing protein data to augment protein inverse folding.

Let $\mathcal{D}=\{\mathcal{P}_r=(\mathcal{S}_r, \mathcal{X}_r)\}_{r=1}^M$ be an external database of $M$ known protein sequences and their structures. Given a query structure $\mathcal{X}$ with an unknown sequence, we introduce a retrieval step to find a set of structurally similar proteins from $\mathcal{D}$. The retrieved set is denoted as $\mathcal{R}$, where 
\begin{equation}
    \mathcal{R} = \{\mathcal{P}_{i_1},\mathcal{P}_{i_2},\cdots\}\subset \mathcal{D}.
\end{equation}
Here, $\mathcal{R}$ contains the proteins most similar to $\mathcal{X}$ in structure based on the designed similarity measurement.

 As shown in Figure~\ref{fig: methods}, the retrieval process contains the following three stages to obtain $\mathcal{R}$: a hierarchical search to search for candidate structures, a residue-wise alignment, and the generation of the amino acid profile.

\subsubsection{Hierarchical Search}
Existing protein databases are generally large-scale and rapidly growing~\citep{varadi2022alphafold}. It is crucial to leverage all available structures as much as possible. However, exhaustive search over all proteins is time-consuming and even impractical.
Therefore, we propose a hierarchical search strategy to efficiently identify proteins with structures similar to the query structure $\mathcal{X}$ from large databases. The hierarchical search first uses a coarse-grained search followed by a fine-grained search.

First, we use FoldSeek~\citep{van2022foldseek} to perform a coarse-grained search of $\mathcal{X}$ against the entire database $\mathcal{D}$. FoldSeek represents 3D structures as sequences of discrete structural alphabet identifiers~(3Di). We leverage $\mathsf{fident}$, defined as the fraction of identical 3Di characters in the alignment between two structures, to perform an initial filtering. We retain only those proteins with a $\mathsf{fident}$ score greater than 0.5. This process yields an initial candidate set $\mathcal{D}'\subset \mathcal{D}$, significantly reducing the search space for the next stage.

Second, we further perform a fine-grained search on $\mathcal{D}'$ using US-align~\citep{zhang2005tm}, which performs coordinate-based structural alignment and yields the TM-score for similarity measurement. Since the TM-score is asymmetric and dependent on the reference protein length, an alignment between two proteins produces two scores $tm_1$ and $tm_2$. For example, we have two proteins with structures “AAA’’ and “AAABBBB’’. When the TM-score is normalized by the length of the shorter protein, it is larger than the TM-score normalized by the length of the longer protein. We take the minimum of $tm_1$ and $tm_2$ so that small proteins that only align with a local region of the query protein can be retained. 
We retain all structures with $\min(tm_1, tm_2)>0.5$, where the threshold is set to 0.5 according to \citep{xu2010significant}.
Finally, the refined protein set is denoted as $\mathcal{R}$. Both Foldseek and US-align are sequence-independent, ensuring that this retrieval process is based solely on structural information $\mathcal{X}$.

\subsubsection{Residue-Wise Alignment}
For each structure $\mathcal{X}_r$ from the retrieved protein $\mathcal{P}_r=(\mathcal{S}_r, \mathcal{X}_r)\in \mathcal{R}$, we construct a residue-wise alignment to the query structure. The purpose of this alignment is to identify locally matched regions, which allows us to use the amino acid types from the retrieved proteins to augment the generation of the query sequence.

The alignment is also produced by the US-align.
As shown in Figure~\ref{fig: methods}, the alignment produces a mapping between residues in the query structure and residues in the retrieved proteins. For each residue position $i$ in the query $\mathcal{X}$, the alignment either identifies a corresponding residue $j$ in $\mathcal{X}_r$ or indicates that position $i$ does not align with any residue.

For each position $i$ in the query sequence, we define a set $\mathcal{T}_i$ to denote the amino acid types of all aligned residues from the retrieved proteins:
\begin{equation}
\mathcal{T}_i = \{ \mathcal{S}_r[j] \mid \forall \mathcal{P}_r=(\mathcal{S}_r, \mathcal{X}_r) \in \mathcal{R} \text{ where residue } i \text{ of } \mathcal{X} \text{ aligns with residue } j \text{ of } \mathcal{X}_r \},
\end{equation}
where $\mathcal{S}_r[j]$ denotes the $j$-th amino acid of sequence $\mathcal{S}_r$. This $\mathcal{T}_i$ reflects the amino acid types observed at structurally aligned positions, providing information about natural sequence variation compatible with the local structural environment of residue $i$.

\subsubsection{Amino Acid Profile Generation}
We generate a position-specific probability matrix, called amino acid profile, from the collected sets $\{\mathcal{T}_i\}_{i=1}^N$.
We denote the amino acid profile as $\bm{\Pi}\in \mathbb{R}^{N\times \vert V_{aa} \vert}$, where $\vert V_{aa}\vert=20$ is the size of the amino acid vocabulary.
For the $i$-th residue and amino acid type $aa\in V_{aa}$, the profile value $\bm{\Pi}_{i, aa}$ is computed as:
\begin{equation}
\label{eqn:AAprofile}
\bm{\Pi}_{i,aa} =
\begin{cases}
\frac{ \text{count}(aa \in \mathcal{T}_i) }{ |\mathcal{T}_i| } & \text{if } |\mathcal{T}_i| > 0 \\
\frac{1}{|V_{aa}|} & \text{if } |\mathcal{T}_i| = 0
\end{cases}
,
\end{equation}
where $\text{count}(aa \in \mathcal{T}_i)$ is the number of times amino acid type $aa$ appears in the set $\mathcal{T}_i$. For the unaligned positions, where no similar structure is retrieved, or no residue on the retrieved structure aligns, the set $\mathcal{T}_i$ will be empty. For these positions, we assign a uniform distribution to provide a non-informative prior.

The resulting profile $\{\bm{\Pi}_{i}\}_{i=0}^{N-1}$ serves as the protein knowledge mined from the natural protein data. This profile is then used as an additional input to our diffusion model, guiding the sequence generation towards amino acid types validated in known structures.

\subsection{Knowledge-Aware Diffusion Model}
\label{diff_model} 
In this section, we introduce our knowledge-aware diffusion model, which consists of the discrete denoising diffusion process and the knowledge-aware guiding modules.
\subsubsection{Discrete Denoising Diffusion}
We follow the discrete denoising diffusion settings in~\citep{austin2021structured}. In our setting, the vocabulary of protein inverse folding contains $K$ kinds of natural amino acids, i.e., $K=\vert V_{aa} \vert=20$. The amino acid feature of sequence $\mathcal{S}$ is denoted as
$\bm{X}^{aa} \in \mathbb{R}^{N \times K}$.
To avoid confusion with coordinate representation denoted by $\bm{x}_i^l$ in Sec~\ref{sec:3d_representation},
we use $\bm{x}^i$ to denote the amino acid feature of the $i$-th amino acid.

\paragraph{Forward Diffusion Process.}
 The forward diffusion process is defined as $q$, which progressively corrupts an initial clean $\bm{X}_0^{\text{aa}}$ over $T$ timesteps. The Markov chain of increasingly noisy sequences is $\bm{X}_0^{aa}, \bm{X}_1^{aa} \ldots, \bm{X}_T^{aa}$. The transition at each step $t$ is defined by a matrix $\bm{Q}_t$, and we add noise to the amino acid type of each node, such that
 \begin{equation}
      q(\bm{X}_t^{aa} \mid \bm{X}_{t-1}^{aa}) = \bm{X}_{t-1}^{aa}\bm{Q}_t.   
 \end{equation}
We use a standard cosine noise schedule $\beta_t$ to define a uniform transition matrix~\citep{nichol2021improved}:
 \begin{equation}
     \bm{Q}_t = (1-\beta_t)\bm{I}+\beta_t\bm{1}_K\bm{1}_K^\top/K, 
 \end{equation} 
where $\bm{1}_K \in \mathbb{R}^{K \times 1}$ denotes the all-one column vector, 
so that $\bm{1}_K \bm{1}_K^\top \in \mathbb{R}^{K \times K}$ is the uniform matrix.
The final state $\bm{X}_T^{aa}$ converges to a uniform distribution over all amino acids and is independent of the input $\bm{X}_0^{aa}$. For any noisy state $\bm{X}_t^{aa}$, it can be sampled in a closed form:
 \begin{equation}
 q(\bm{X}_t^{aa} \mid \bm{X}_0^{aa}) = \bm{X}_0^{aa} \bar{\bm{Q}}_t, 
\end{equation}
where $\bar{\bm{Q}}_t = \prod_{k=1}^t \bm{Q}_k$.
\paragraph{Training Objective.}
The model is trained to predict the original clean sequence $\bm{X}_0^{aa}$ given the noisy sequence $\bm{X}_t^{aa}$ at timestep $t$, along with the condition including protein structure $\mathcal{X}$ and the retrieved similar structures $\mathcal{R}$.
The objective $\mathcal{L}$ is to minimize the cross-entropy loss between its prediction and the true clean sequence:
\begin{equation}
\mathcal{L} = \mathbb{E}_{t, \bm{X}_0^{aa}, \bm{X}_t^{aa}} 
\left[ L_{\text{CE}}\big(\mathcal{F}_\theta(\bm{X}_t^{aa}, t, \mathcal{X}, \mathcal{R}), \bm{X}_0^{aa} \big) \right],
\end{equation}
where $\mathcal{F}_\theta$ denotes a network $\mathcal{F}$ with parameter $\theta$ and $L_{\text{CE}}$ denotes the cross-entropy loss.
\paragraph{Reverse Denoising Process.}
The reverse process is defined as $p_\theta(\bm{X}_{t-1}^{aa} \mid \bm{X}_t^{aa})$, which aims to denoise a sequence from $\bm{X}_T^{aa}$ back to a clean sequence $\bm{X}_0^{aa}$. 
The generative distribution $p_\theta(\bm{X}_{t-1}^{aa} \mid \bm{X}_t^{aa})$
is parameterized using the distribution
$\hat{p}_\theta(\bm{\hat{x}}_0^i \mid \bm{x}_t^i)$,
where $\hat{p}_\theta(\bm{\hat{x}}_0^i \mid \bm{x}_t^i)$ represents the predicted
distribution over amino-acid types for the $i$-th residue  by the trained denoising network.
Specifically, we marginalize over the network's predictions for the clean sequence to compute the distribution for each residue $i$~\citep{yi2023graph}:
\begin{equation}
 p_\theta(\bm{x}_{t-1}^i \mid \bm{x}_t^i) \propto \sum_{\bm{\hat{x}}_0^i} q(\bm{x}_{t-1}^i \mid \bm{x}_t^i, \bm{\hat{x}}_0^i) \hat{p}_\theta(\bm{\hat{x}}_0^i \mid \bm{x}_t^i).
\end{equation}
The posterior distribution $q(\bm{x}_{t-1}^i \mid \bm{x}_t^i, \bm{\hat{x}}_0^i)$ can be calculated in closed form using Bayes' theorem:
\begin{align}
q(\bm{x}_{t-1}^i \mid \bm{x}_t^i, \bm{\hat{x}}_0^i) &=\frac{q(\bm{x}_t^i \mid \bm{x}_{t-1}^i,\bm{\hat{x}}_0^i) q(\bm{x}_{t-1}^i \mid \bm{\hat{x}}_0^i)}{q(\bm{x}_t^i \mid \bm{\hat{x}}_0^i)} \\
&= \text{Cat}\left(\bm{x}_{t-1}^i;\frac{\bm{x}_t^i\bm{Q}_t^\top\odot\bm{\hat{x}}_0^i\bar{\bm{Q}}_{t-1}}{\bm{\hat{x}}_0^i \bar{\bm{Q}_t} {\bm{x}_t^i}^{\top}}\right) ,
\end{align}
where $\text{Cat}(\bm{x}^i;\cdot)$ is a categorical distribution over $\bm{x}^i$. The probability for the entire sequence is the product of the individual amino acid probabilities:
\begin{equation}
 p_\theta(\bm{X}_{t-1}^{aa} \mid \bm{X}_t^{aa}) = \prod_{1\le i \le N} p_\theta(\bm{x}_{t-1}^i \mid \bm{x}_t^i).   
\end{equation}
To generate a completely new sequence, the process begins with a random noise sequence sampled from  $\bm{X}_T^{aa}$. This sequence is then iteratively denoised at each timestep using the reverse denoising process, eventually converging to a clean sequence $\bm{X}_0^{aa}$. To accelerate this iterative generation, we employ a discrete denoising diffusion implicit model~(DDIM)~\citep{song2020denoising} sampler~\citep{bai2025mask}.

\subsubsection{Knowledge-aware Guiding Modules}
We design two knowledge-aware guiding modules for the generation process, which consists of the profile integration module and the masked sequence designer~(MSD)~\citep{bai2025mask} module. The profile integration module captures knowledge from structurally similar proteins, and the MSD module learns knowledge from protein sequence.

\paragraph{Profile Integration.}
The retrieval-based amino acid profile $\bm{\Pi}_{i}$ is integrated with protein structure representations $\bm{h}^L_i$ to guide the diffusion model. 
The integration of $\bm{\Pi}_{i}$ and $\bm{h}^L_i$ is achieved via a lightweight module. The profile vector $\bm{\Pi}_{i}$ is projected to match the hidden dimension of the node features and then integrated with the final node representation $\bm{h}_i^L$ via a residual connection. The resulting feature is further updated through a MLP and the softmax function is used to compute the probability over amino acid types:
\begin{align}
\bm{z}_i &= \phi_2(\phi_1(\bm{\Pi}_{i}) + \bm{h}_i^L),\\
\textbf{prob}_i &= \text{softmax}(\bm{z}_i),
\label{p_i}
\end{align}
where $\phi_1$ and $\phi_2$ are MLPs. 

\paragraph{Masked Sequence Designer.}
To incorporate prior knowledge from protein sequence, we follow MapDiff~\citep{bai2025mask} by pre-training a separate MSD. The role of MSD is to refine residues with low predictive confidence during the denoising process.

The backbone of the MSD is an invariant point attention~(IPA) network, which is initially proposed by AlphaFold2~\citep{jumper2021highly} and later modified by Frame2seq~\citep{akpinaroglu2023structure} to integrate geometric information. We use a masked language modeling (MLM) objective~\citep{devlin2019bert} to learn the natural amino acid distribution. Specifically, for each training sequence, a portion of the amino acids is randomly corrupted. Among them, 80\% are replaced with a special [MASK] token, 10\% are replaced with random amino acids, and the remaining 10\% remain unchanged.
The MSD takes the masked sequence and the corresponding backbone coordinates as input and is trained with a cross-entropy loss to predict the original amino acid types. The MSD is pre-trained first and then frozen during the training of the whole RadDiff.

To quantify the confidence of the prediction, we define the entropy of the predicted residue $i$ as:
\begin{equation}
\text{ent}_i = -\sum_j \textbf{prob}_{ij} \log(\textbf{prob}_{ij}),
\label{ent_i}
\end{equation}
where $\textbf{prob}_{ij}$ represents the probability of amino acid type $j$ for the $i$-th residue. 
The amino acids with the lowest entropy are masked and re-predicted by the MSD. We use annotations with superscript $m$ to denote the output by the MSD. The output representation for residue $i$ generated by MSD is denoted as $\bm{z}_i^m$, and the probability distribution is computed as $\textbf{prob}_i^m = \text{softmax}(\bm{z}_i^m)$.
Similarly, we define the entropy of the re-predicted residue $i$ as:
\begin{align}
\text{ent}_i^m = -\sum_j \textbf{prob}_{ij}^m \log(\textbf{prob}_{ij}^m).
\end{align}

Finally, the final probability $\textbf{prob}_i^f$ for amino acid $i$ is computed as:
\begin{equation}
\textbf{prob}_i^f = \text{softmax} \left( \frac{\exp(-\text{ent}_i)}{\exp(-\text{ent}_i) + \exp(-\text{ent}_i^m)} \bm{z}_i + \frac{\exp(-\text{ent}_i^m)}{\exp(-\text{ent}_i) + \exp(-\text{ent}_i^m)} \bm{z}_i^m \right).
\end{equation}

This refinement process ensures that the final prediction is influenced by both the original and the re-predicted probability, yielding more accurate and confident amino acid type prediction.

\section{Experiment}
\subsection{Evaluation Settings}
\noindent\textbf{Database for Retrieval and Data Leakage Prevention.}
\begin{figure}
\centering
\includegraphics[width=0.3\textwidth]{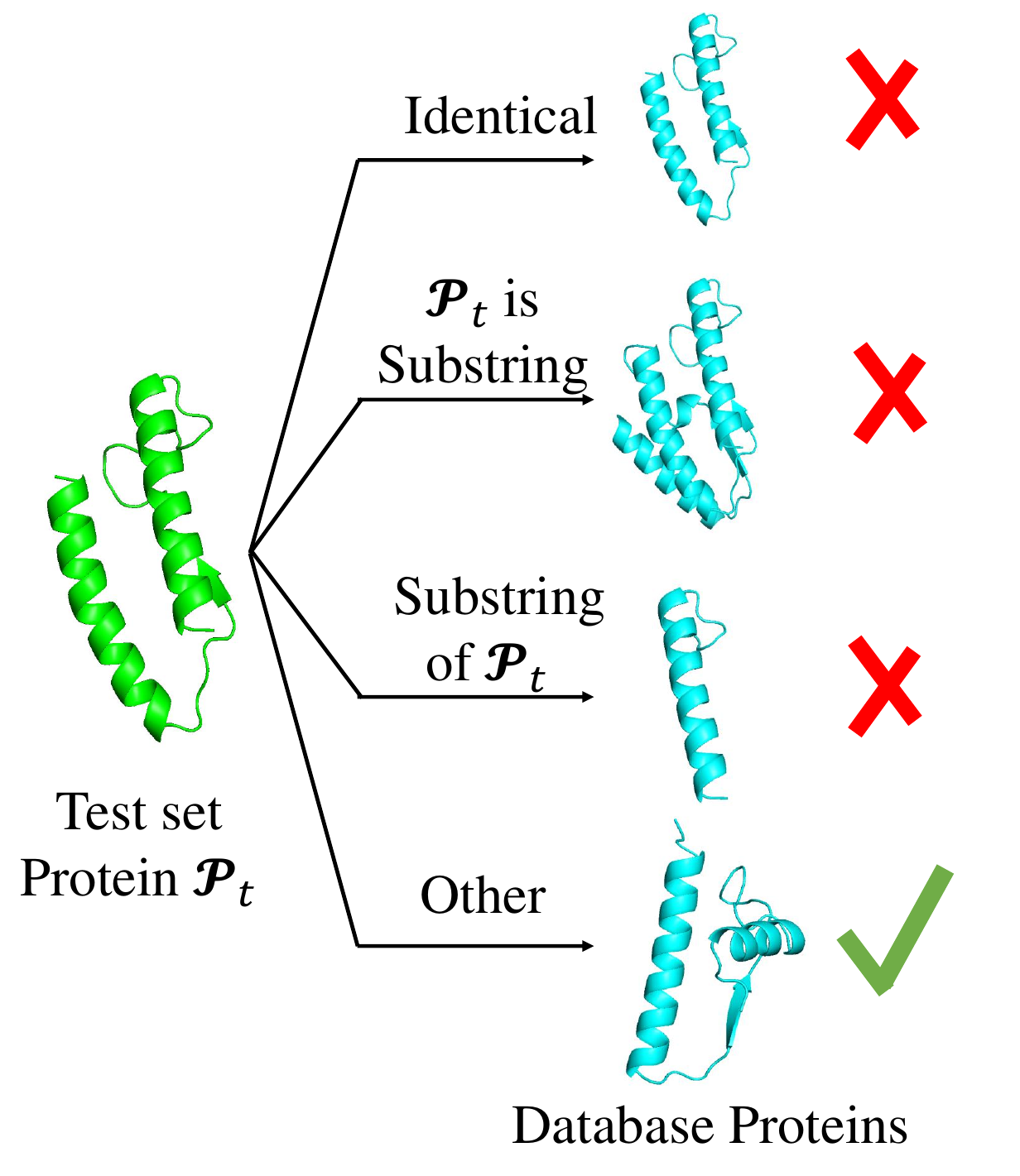}
\caption{Illustration of strategy to prevent data leakage.}
\label{fig: db_filter}
\end{figure}
We utilize the AlphaFold predicted Swiss-Prot database~\citep{varadi2022alphafold,bairoch1997swiss}, which contains 542,380 protein structures, as the external protein database for retrieval. As shown in Figure~\ref{fig: db_filter}, we implement a strict filtering strategy containing identity filtering and substring filtering during the retrieval process to prevent data leakage. 
The identity filtering means that if the structure of the retrieved protein is identical to the query structure, it will be discarded.
The substring filtering prevents data leakage from the sequence level. For domain-based datasets like CATH, where the samples may be cropped from the proteins in the database, the substring filtering means that if the amino acid sequence of the retrieved protein contains the query's sequence as a substring, or if the query's sequence contains the sequence of the retrieved protein as a substring, the retrieved protein will be discarded.
This filtering strategy ensures that the retrieval augmentation is derived from truly homologous structures rather than from artifacts of dataset construction.

\noindent\textbf{Baselines.}
We adopt two categories of baseline methods for comparison: structure-only methods and knowledge-based methods.
The structure-only methods can be further divided into GNN-based methods and diffusion-based methods. GNN-based methods include AlphaDesign~\citep{gao2022alphadesign}, ProteinMPNN~\citep{dauparas2022robust}, StructGNN~\citep{ingraham2019generative}, GraphTrans~\citep{ingraham2019generative}, GVP~\citep{jinglearning}, and PiFold~\citep{gaopifold}. Diffusion-based methods include GradeIf~\citep{yi2023graph} and MapDiff~\citep{bai2025mask}. Knowledge-based methods include LM-Design~\citep{zheng2023structure}, KW-design~\citep{gaokw} and PRISM~\citep{mahbub2025prism}. All baselines are evaluated under identical experimental settings.

\noindent\textbf{Datasets.}
We adopt several standard benchmark datasets to evaluate model performance.
The CATH v4.2 and v4.3 are two widely used datasets~\citep{orengo1997cath}. By following previous methods, a topology-based data split is performed to prevent overlap between the training, validation, and test sets. The CATH v4.2 dataset is split into 18,024 training, 608 validation, and 1,120 test samples. The CATH v4.3 dataset is split into 16,630 training, 1,516 validation, and 1,864 test samples.
To further evaluate the zero-shot generalization capability of the methods, we include two independent test sets, TS50~\citep{li2014direct} and PBD2022~\citep{berman2000protein}. TS50~\citep{li2014direct} is a widely-used benchmark containing 50 diverse protein chains. PDB2022 dataset, curated by ~\citep{zhou2023prorefiner}, consists of 1,975 structures published in the Protein Data Bank~(PDB)~\citep{berman2000protein} between January 5, 2022, and October 26, 2022, which is after the release date of CATH. Both datasets are disjoint from the CATH training set, which can be used to evaluate the structural and temporal generalization of baselines~\citep{bai2025mask}.

\subsection{Implementation Details}
The denoising network's backbone consists of EGNN with 6 layers, each with a hidden dimension of 128. The masked sequence designer is composed of 6 IPA layers, also with a hidden dimension of 128.
We employ the Adam optimizer with an initial learning rate of $5\times 10^{-4}$, managed by a one-cycle learning rate scheduler. A batch size of 8 is used for all training stages. Following the protocol of ~\citep{bai2025mask}, the MSD module is pre-trained for 200 epochs. The main graph-based denoising model is trained for 100 epochs.
For the retrieval process, FoldSeek~\citep{van2022foldseek} of version \texttt{9.427df8a} is used.  
For the \textit{in silico} foldability analysis, we use Boltz v2.03~\citep{passaro2025boltz2} to predict the 3D structures of the generated sequences. The MSAs, which are required as input for Boltz, are generated using the online MSA server.

\subsection{Protein Design on CATH}
\noindent\textbf{Accuracy}. We compare the perplexity and sequence recovery rate of RadDiff and baselines on the CATH v4.2 and CATH v4.3 datasets. Perplexity measures the negative log-likelihood of the native amino acid sequence under the predicted probability distributions. Sequence recovery rate measures the fraction of the predicted amino acids that match the native sequence. As shown in Table~\ref{tab:cath_results}, RadDiff consistently outperforms all baseline methods in both perplexity and sequence recovery rate across the short, single-chain, and full dataset splits to achieve a new state-of-the-art.
Specifically, on the full CATH v4.2 dataset, RadDiff achieves a perplexity of 2.46 and a sequence recovery rate of 67.14\%. This corresponds to a 9.23\% reduction in perplexity and a 10.01\% relative improvement in recovery rate over the the existing methods. On the short and single-chain subsets of CATH v4.2, RadDiff improves recovery rate by 19.91\% and 9.59\%, respectively.
RadDiff shows stronger performance on the full CATH v4.3 dataset, achieving a perplexity of 2.48 and sequence recovery rate of 72.40\%. This corresponds to a 39.0\% reduction in perplexity and a 19.0\% improvement in recovery rate compared to the previous best method. On the short and single-chain subsets of CATH v4.3, RadDiff improves recovery rate by 35.16\% and 37.24\%, respectively.
Overall, the experimental results show that RadDiff effectively leverages retrieval-augmented knowledge to reduce perplexity and improve sequence recovery rate in prediction.

\begin{table*}[t]
\centering
\caption{Performance on CATH v4.2 and CATH v4.3 datasets.}
\label{tab:cath_results}
\scalebox{0.75}{
\begin{tabular}{lccccccc}
\toprule[0.8pt]
\multirow{2}{*}{Models} & \multirow{2}{*}{Model Size} & \multicolumn{3}{c}{Perplexity ($\downarrow$)} & \multicolumn{3}{c}{Median Recovery Rate (\%, $\uparrow$)} \\
\cmidrule(lr){3-5} \cmidrule(lr){6-8}
& & Short & Single-chain & Full & Short & Single-chain & Full \\
\hline
\multicolumn{8}{c}{\textbf{CATH v4.2}} \\
\hline
StructGNN~\citep{ingraham2019generative} &  1.4M & 8.29 & 8.74 & 6.40 & 29.44 & 28.26 & 35.91 \\
GraphTrans~\citep{ingraham2019generative} &  1.5M & 8.39 & 8.83 & 6.63 & 28.14 & 28.46 & 35.82 \\
GVP~\citep{jinglearning} &  2.0M & 7.09 & 7.49 & 6.05 & 32.62 & 31.10 & 37.64 \\
AlphaDesign~\citep{gao2022alphadesign} &  6.6M & 7.32 & 7.63 & 6.30 & 34.16 & 32.66 & 41.31 \\
ProteinMPNN~\citep{dauparas2022robust} &  1.9M & 6.90 & 7.03 & 4.70 & 36.45 & 35.29 & 48.63 \\
PiFold~\citep{gaopifold}  & 6.6M & 5.97 & 6.13 & 4.61 & 39.17 & 42.43 & 51.40 \\
LM-Design~\citep{zheng2023structure}  & 659M & 6.86 & 6.82 & 4.55 & 37.66 & 38.94 & 53.19 \\
KW-Design~\citep{gaokw} &798M & 5.48 & 5.16 & 3.46 & 44.66 &45.45 & 60.77\\
PRISM~\citep{mahbub2025prism}&-&3.74&2.68&2.71&40.98&60.89&60.43\\
GradeIf~\citep{yi2023graph}  & 7.0M & 5.65 & 6.46 & 4.40 & 45.84 & 42.73 & 52.63 \\
MapDiff~\citep{bai2025mask}  & 14.1M & 3.99 & 4.43 & 3.46 & 52.85 & 50.00 & 61.03 \\
RadDiff      &14.2M  & \textbf{2.97} & \textbf{2.55} & \textbf{2.46} & \textbf{63.37} & \textbf{66.73} & \textbf{67.14} \\
\midrule 
\multicolumn{8}{c}{\textbf{CATH v4.3}} \\
\hline
GVP-GNN-Large~\citep{hsu2022learning}  & 21M & 7.68 & 6.12 & 6.17 & 32.60 & 39.40 & 39.20 \\
ProteinMPNN~\citep{dauparas2022robust}  & 1.9M & 6.12 & 6.18 & 4.63 & 40.00 & 39.13 & 47.66 \\
PiFold~\citep{gaopifold}  & 6.6M & 5.52 & 5.00 & 4.38 & 43.06 & 45.54 & 51.45 \\
LM-Design~\citep{zheng2023structure}  & 659M & 6.01 & 5.73 & 4.47 & 44.44 & 45.31 & 53.66 \\
KW-Design~\citep{gaokw} &798M & 5.47 & 5.23 & 3.49 & 43.86 &45.95 & 60.38\\
GradeIf~\citep{yi2023graph}  & 7.0M & 5.30 & 6.05 & 4.58 & 48.21 & 45.94 & 52.24 \\
MapDiff~\citep{bai2025mask}  & 14.1M & 3.88 & 3.85 & 3.48 & 55.95 & 54.65 & 60.86 \\
RadDiff  &14.2M  & \textbf{2.48} & \textbf{2.35} & \textbf{2.38} & \textbf{75.62} & \textbf{75.00} & \textbf{72.40} \\
\bottomrule[0.8pt]
\end{tabular}
}
\end{table*}
\noindent\textbf{Model Size.} 
We also compare the model size of RadDiff and baselines. Since PRISM is not open-sourced, the model size information is unavailable. As shown in Table~\ref{tab:cath_results}, a key advantage of RadDiff is its parameter efficiency compared with the other knowledge-based methods. While PLM-based methods like LM-Design and KW-Design also leverage external protein knowledge, they have high parameter overhead, requiring $46\times$ and $56\times$
 more parameters than RadDiff, respectively. In contrast, RadDiff successfully integrates the protein knowledge while keeping a lightweight model architecture.

\begin{table*}[t]
\centering
\caption{Generalizability evaluation on PDB2022 and TS50 datasets. The results in brackets are from the model trained with CATH v4.3.}
\label{tab:genelization_results}
\scalebox{0.7}{
\begin{tabular}{lcccccc}
\toprule[0.8pt]
\multirow{2}{*}{Models} & \multicolumn{3}{c}{PDB2022} & \multicolumn{3}{c}{TS50} \\
\cmidrule(lr){2-4} \cmidrule(lr){5-7}
 & Recovery($\uparrow$) & NSSR62($\uparrow$) & NSSR90($\uparrow$) & Recovery($\uparrow$) & NSSR62($\uparrow$) & NSSR90($\uparrow$) \\
\midrule
ProteinMPNN~\citep{dauparas2022robust} & 56.75 (56.65) & 72.50 (72.59) & 69.96 (69.95) & 52.34 (51.80) & 70.31 (70.13) & 66.77 (66.80) \\
PiFold~\citep{gaopifold}      & 60.63 (60.26) & 75.55 (75.30) & 72.96 (72.86) & 58.39 (58.90) & 73.55 (74.52) & 70.33 (71.33) \\
LM-Design~\citep{zheng2023structure}   & 66.03 (66.20) & 79.55 (80.12) & 77.60 (78.20) & 57.62 (58.27) & 73.74 (75.69) & 71.22 (73.12) \\
GradeIf~\citep{yi2023graph}    & 58.09 (58.35) & 77.44 (77.51) & 74.57 (74.97) & 57.74 (59.27) & 77.77 (79.11) & 74.36 (76.24) \\
MapDiff~\citep{bai2025mask}     & 68.03 (68.00) & 84.19 (84.30) & 82.13 (82.29) & 68.76 (69.77) & 84.10 (85.27) & 81.76 (83.08) \\
RadDiff        & \textbf{76.22 (75.70)} & \textbf{87.38 (85.62)} & \textbf{86.37 (84.06)} & \textbf{75.64 (76.99)} & \textbf{88.98 (91.10)} & \textbf{86.91 (88.65)} \\   
\bottomrule[0.8pt]
\end{tabular}
}
\end{table*}

\subsection{Zero-shot Generalization on TS50 and PDB2022}
We evaluate the zero-shot performance of RadDiff and baselines on two independent datasets, TS50 and PDB2022, using models trained on CATH v4.2 and CATH v4.3. The evaluation metrics include sequence recovery rate and native sequence similarity recovery~(NSSR)~\citep{loffler2017rosetta}. NSSR measures the biochemical similarity between predicted and native residues using the BLOSUM~\citep{henikoff1992amino} substitution matrix, where a residue pair is considered as a match if its BLOSUM score is positive. NSSR62 and NSSR90 denote the use of the BLOSUM62 and BLOSUM90 matrices, respectively.
As shown in Table~\ref{tab:genelization_results}, we can find that RadDiff consistently outperforms all baselines in both recovery rate and NSSR, regardless of the training dataset.
On the PDB2022 dataset, models trained on CATH v4.2 and v4.3 achieve recovery rate of 76.22\% and 75.70\% respectively, improving over the previous best methods by 12.04\% and 11.32\%.
On the TS50 dataset, the models achieve recovery rate of 75.64\% and 76.99\% respectively, improving over the previous best methods by 10.00\% and 10.35\%.
Furthermore, RadDiff obtains the highest NSSR62 and NSSR90 scores, demonstrating its superior ability to not only predict the correct amino acid but also to capture biochemically meaningful residue similarities. Overall, the results show that RadDiff generalizes well on unseen data.

\begin{table}[t]
\centering
\caption{Foldability comparison using Boltz and ESMFold. }
\label{tab:design_methods}
\scalebox{0.7}{
\begin{tabular}{lcccccccc}
\toprule
\multirow{2}{*}{Models} & \multicolumn{4}{c}{Boltz2} & \multicolumn{2}{c}{ESMFold} \\
\cmidrule(lr){2-5} \cmidrule(lr){6-7}
 & TMscore ($\uparrow$) & RMSD ($\downarrow$) & pTM ($\uparrow$) & pLDDT ($\uparrow$) & TMscore ($\uparrow$) & RMSD ($\downarrow$) \\
\midrule
ProteinMPNN~\citep{dauparas2022robust}& 84.95 \std{16.36}  &1.66 \std{0.94} & 82.66 \std{15.47} & 86.74 \std{10.73} &84.34 \std{18.00} &$\bm{1.78}$ \std{\textbf{1.03}}\\
PiFold~\citep{gaopifold}     & 84.77 \std{15.85} & 1.72 \std{0.90} & 82.48 \std{14.33} & 86.08 \std{10.04} & 81.82 \std{18.63} & 1.97 \std{1.10} \\
LM-Design~\citep{zheng2023structure}  & 83.98 \std{16.78} & 1.73 \std{0.94} & 83.11 \std{14.67} & 87.16 \std{9.69} & 80.87 \std{19.33} & 1.98 \std{1.13} \\
GradeIf~\citep{yi2023graph}   & 78.55 \std{17.46} & 2.27 \std{0.97} & 74.43 \std{15.04} & 78.50 \std{11.50} & 73.79 \std{20.94} & 2.58 \std{1.26} \\
MapDiff~\citep{bai2025mask}    & 84.71 \std{14.94} & 1.78 \std{0.83} & 82.32 \std{12.72} & 86.04 \std{8.86} & 82.03 \std{17.00} & 2.01 \std{1.00} \\
RadDiff & 
$\bm{87.69}$ \std{\textbf{13.06}} & $\bm{1.55}$ \std{\textbf{0.76}} & 
$\bm{85.58}$ \std{\textbf{11.32}} & $\bm{89.70}$ \std{\textbf{6.78}} & 
$\bm{85.43}$ \std{\textbf{14.74}} & 1.79\std{0.90} \\
\bottomrule
\end{tabular}
}
\end{table}

\subsection{Foldability}
We also compare the foldability of RadDiff and baselines.
In addition to sequence recovery rate, \textit{in silico} foldability is critical to measure if the generated sequence will fold into the intended structure. To evaluate the foldability, we employ two cutting-edge structure prediction models, the MSA-based method Boltz2~\citep{passaro2025boltz2} and MSA-free method ESMFold~\citep{lin2022language}, to predict the 3D structures of sequences. AlphaFold2 is not used due to its prohibitively expensive local MSA search process. 
As shown in Table~\ref{tab:design_methods}, we compare the re-folded structures to the ground-truth crystal structures using a suite of metrics, including predicted TM-score~(pTM), predicted aligned error~(PAE), and predicted local distance difference test~(pLDDT) from Boltz2, as well as the TM-score and RMSD from direct structural alignment. 
The comparison is focused specifically on samples where a protein hit is successfully retrieved from the database, allowing for a direct comparison of design quality when our method is successfully guided by retrieved proteins. 
In the Boltz2 prediction results, RadDiff shows higher structural similarity to the native fold and achieves higher confidence scores across all metrics. For the ESMFold predictions, RadDiff achieves the best TM-score compared with the other methods and is comparable to ProteinMPNN in RMSD. 
Overall, the results demonstrate that RadDiff produces designs that are more likely to fold into the intended structures than baselines.

\subsection{Analysis of Retrieval Augmentation}
\noindent\textbf{Retrieval Time.}
We evaluate the retrieval time of the retrieval augmentation process of RadDiff. To quantify the retrieval time, we calculate the runtime of our hierarchical search strategy for all 1,120 query proteins in the CATH v4.2 test set against the Swiss-Prot database which contains 542,380 structures. This corresponds to a total search space of over 600 million pairwise comparisons.
As shown in Table~\ref{retrieval_time}, the results demonstrate the efficiency of the retrieval process. The entire retrieval process for all 600 million pairs is completed in just 306.5 seconds, corresponding to an average of only 0.27 seconds per query.
The initial rapid filtering with FoldSeek requires only 54.0 seconds in total, corresponding to an average runtime of 0.04 seconds per query. Subsequently, the more time-consuming US-align step, which provides the high-quality alignments essential for our method, requires only a few minutes. The results demonstrate that RadDiff is efficient and computationally practical. 

\noindent\textbf{Impact of Retrieval Augmentation.} 
We show the perplexity and recovery rate of samples where a protein hit is successfully retrieved from the database.
We divide the samples in the test set into two subsets, ``w. RAG'' and ``w.o. RAG''. The ``w. RAG'' subset contains proteins that have at least one hit in the database, and the ``w.o. RAG'' subset contains proteins that fail to find any similar structures in the database.
The experiment is conducted on the CATH v4.2 dataset.
As shown in Table~\ref{tab: RAG_data}, the ratio of the ``w. RAG'' subset is 47.86\% and ``w.o. RAG'' subset is 52.14\%.
RadDiff achieves 89.80\% recovery rate on the ``w. RAG'' subset, which is 31\% higher than the result in ``w.o. RAG'' subset. In addition, the predictions also show substantially higher confidence when the retrieved structures are available, with a perplexity of 1.56 on the ``w. RAG'' subset compared to 4.01 on the ``w.o. RAG'' subset. 
These results demonstrate that integrating knowledge from the protein database provides strong guidance to the generative process and significantly boosts performance. Nevertheless, even when no structural hits are available, RadDiff still produces designs with more than 58\% recovery rate.

\begin{figure}[ht]
    \center
    \includegraphics[scale=0.40]{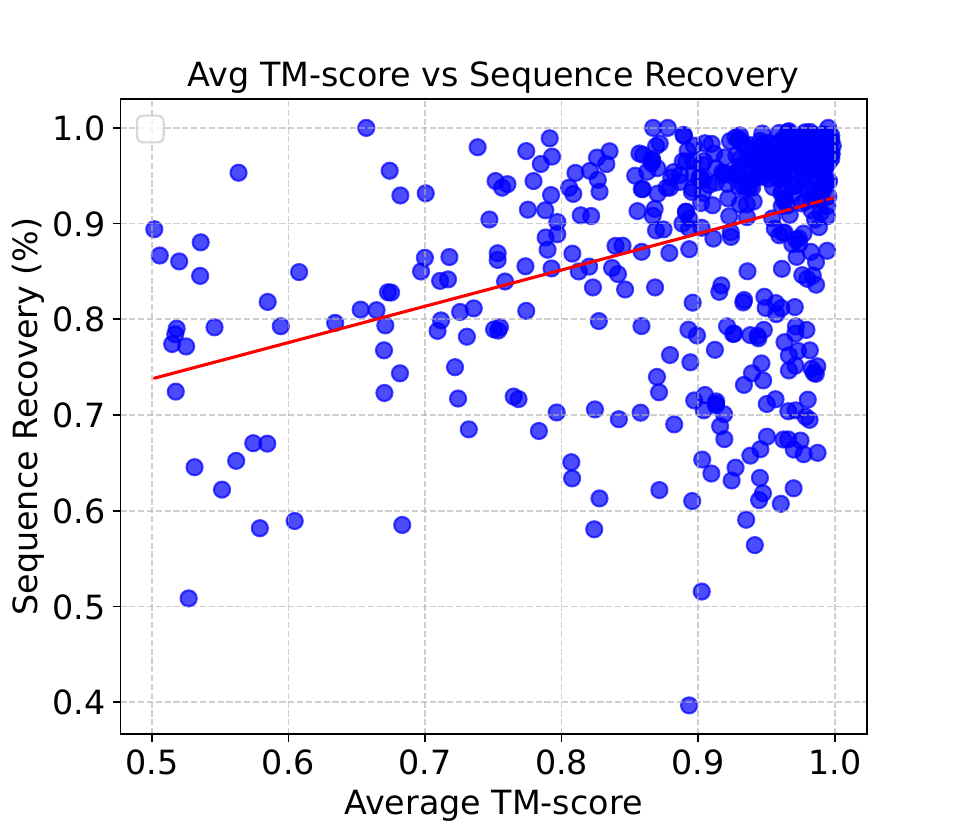}
    \caption{The relationship between the average TM-score of test proteins and sequence recovery rate.}    
    \label{fig_results: b}
\end{figure}

\noindent\textbf{Influence of Structure Similarity.}  
We investigate how the structural similarity between the test set and the retrieved proteins will influence the sequence recovery rate.
For each test sample in the ``w. RAG'' subset, we calculated the average TM-score across all of its retrieved structures and show this value against the recovery rate.
As shown in Figure~\ref{fig_results: b}, the results show a positive correlation of the TM-score and the recovery rate. The Pearson correlation coefficient is 0.374. This indicates that retrieving more structurally similar proteins generally leads to higher recovery rate.
Moreover, the results also highlight the robustness of RadDiff. Even when the average structural similarity of the retrieved set is modest (e.g., TM-score between 0.5 and 0.7), RadDiff consistently generates sequence with high recovery rate, often exceeding 70-80\%.
\begin{table*}[t]
\centering
\begin{minipage}{0.50\textwidth}
    \centering
    \caption{Retrieval time for searching the CATH v4.2 dataset against the SwissProt database. }    
    \label{retrieval_time}
    \scalebox{0.90}{
    \begin{tabular}{c|ccc}
    \toprule        
        &FoldSeek    & US-align &Total\\
    \midrule
    Time &  53.98s &252.52s &306.5s\\
    Time per Query &0.04s&0.23s&0.27s\\
    \bottomrule
  \end{tabular}
    }
\end{minipage}
\hfill
\begin{minipage}{0.48\textwidth}
    \centering
    \caption{Performance comparison between the ``w. RAG'' and ``w.o. RAG'' subsets.}    
    \label{tab: RAG_data}
    \scalebox{0.8}{
    \begin{tabular}{lcc}
    \toprule
    Metric & w. RAG & w.o. RAG \\
    \midrule
    Ratio(\%) &47.86 &52.14\\
    Recovery Rate (\%)$\uparrow$ & 89.80 & 58.64 \\
    Perplexity$\downarrow$  & 1.56  & 4.01 \\
    \bottomrule
    \end{tabular}
    }
\end{minipage}
\end{table*}

\subsection{Ablation Study}
We conduct an ablation study on the retrieval augmentation module and the MSD module to systematically evaluate the contributions of the key components. We evaluate three distinct model configurations: (1) the full RadDiff model with both modules enabled, (2) a variant with only the retrieval augmentation module, and (3) a variant with only the MSD module.
As shown in Table~\ref{ablation}, we can find that the retrieval augmentation module contributes significantly to the performance, which increases the sequence recovery rate by 6.64\% and reduces perplexity by 0.81. This confirms that the protein knowledge captured by retrieval augmentation is effective for protein inverse folding. The MSD module also contributes positively, improving recovery by 4.13\% and reducing perplexity by 0.93, indicating that MSD successfully enhances prediction confidence by integrating knowledge from protein sequences. Overall, the ablation study demonstrates the effectiveness of the retrieval augmentation and MSD module.
\begin{table}[ht]
\caption{Ablation study.}
  \label{ablation}
  \centering
  \begin{tabular}{cc|cc}
    \toprule        
    MSD&    retrieval augmentation    & Perplexity($\downarrow$) & Recovery Rate(\%, $\uparrow$)   \\
    \midrule
    \Checkmark & \XSolidBrush   &3.27     & 61.50   \\
    \XSolidBrush & \Checkmark   & 3.39    & 63.03     \\
    \midrule
    \Checkmark & \Checkmark   & \textbf{2.46}  & \textbf{ 67.14}    \\
    \bottomrule
  \end{tabular}
\end{table}
\begin{figure}
\centering
\includegraphics[width=0.50\textwidth]{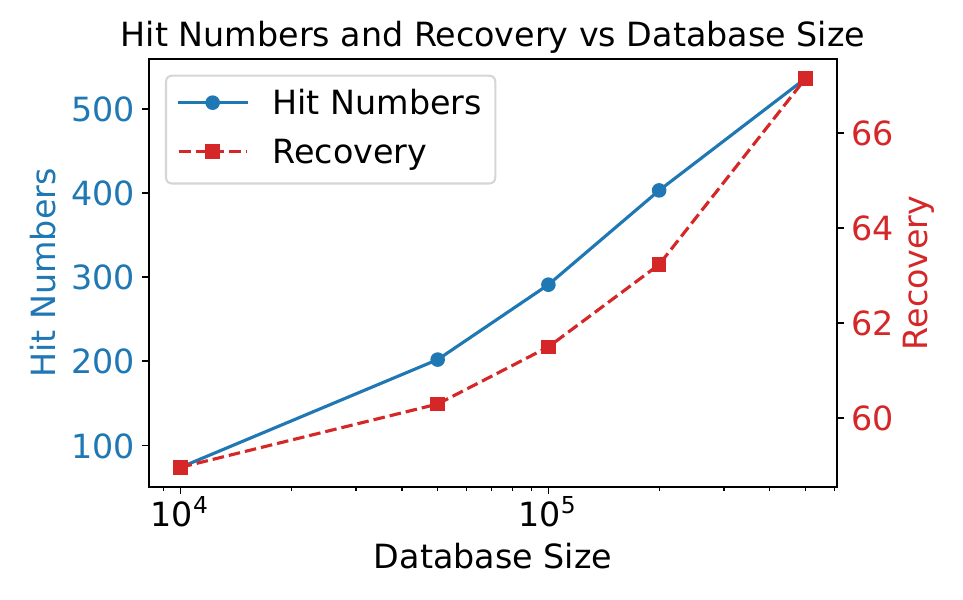}
\caption{The relationship between the size of the protein database, hit numbers, and sequence recovery. }
\label{fig: database_size}
\end{figure}

\subsection{Influence of Database Size}
We investigate how the size of the database will influence the model performance. We vary the size of the database and calculate the retrieval hit numbers and the sequence recovery rate. The hit numbers represent the number of test-set proteins that have structurally similar hits in the database.
As shown in Figure~\ref{fig: database_size}, the hit numbers increase consistently with larger database sizes. This increase directly translates to improved sequence recovery rate. Specifically, when the database contains only 10,000 proteins, the sequence recovery rate is below 60\%. As the database size grows, the recovery rate continues to increase, surpassing 67\%.
The result shows the potential of increasing the database size to improve the performance of RadDiff.


\section{Conclusion}
In this paper, we propose a novel retrieval-augmented denoising diffusion method, called RadDiff, for protein inverse folding. In RadDiff, a novel retrieval-augmentation mechanism is designed to first retrieve a set of structurally similar proteins through hierarchical search, and then perform residue-wise alignment to identify matched structural regions. From the aligned residues, RadDiff constructs a position-specific amino acid profile that captures up-to-date protein knowledge.
RadDiff also designs a knowledge-aware diffusion model to integrate protein knowledge from the amino acid profile through a lightweight module, making RadDiff more parameter-efficient than PLM-based methods. 
Experimental results show that RadDiff consistently outperforms existing methods.

\newpage
\bibliography{iclr2026_conference}

@article{zhang2005tm,
  title={TM-align: a protein structure alignment algorithm based on the TM-score},
  author={Zhang, Yang and Skolnick, Jeffrey},
  journal={Nucleic Acids Research},
  volume={33},
  number={7},
  pages={2302--2309},
  year={2005},  
}

@inproceedings{fu2022sipf,
  title={Sipf: Sampling method for inverse protein folding},
  author={Fu, Tianfan and Sun, Jimeng},
  booktitle={Conference on Knowledge Discovery and Data Mining},
  year={2022}
}

@article{zhang2022us,
  title={US-align: universal structure alignments of proteins, nucleic acids, and macromolecular complexes},
  author={Zhang, Chengxin and Shine, Morgan and Pyle, Anna Marie and Zhang, Yang},
  journal={Nature Methods},
  volume={19},
  number={9},
  pages={1109--1115},
  year={2022},  
}

@article{zhang2004scoring,
  title={Scoring function for automated assessment of protein structure template quality},
  author={Zhang, Yang and Skolnick, Jeffrey},
  journal={Proteins: Structure, Function, and Bioinformatics},
  volume={57},
  number={4},
  pages={702--710},
  year={2004},  
}

@article{xu2010significant,
  title={How significant is a protein structure similarity with TM-score= 0.5?},
  author={Xu, Jinrui and Zhang, Yang},
  journal={Bioinformatics},
  volume={26},
  number={7},
  pages={889--895},
  year={2010},  
}

@article{van2022foldseek,
  title={Foldseek: fast and accurate protein structure search},
  author={van Kempen, Michel and Kim, Stephanie S and Tumescheit, Charlotte and Mirdita, Milot and Gilchrist, Cameron LM and S{\"o}ding, Johannes and Steinegger, Martin},
  journal={biorxiv},  
  year={2022},  
}

@article{steinegger2017mmseqs2,
  title={MMseqs2 enables sensitive protein sequence searching for the analysis of massive data sets},
  author={Steinegger, Martin and S{\"o}ding, Johannes},
  journal={Nature Biotechnology},
  volume={35},
  number={11},
  pages={1026--1028},
  year={2017},  
}

@article{alford2017rosetta,
  title={The Rosetta all-atom energy function for macromolecular modeling and design},
  author={Alford, Rebecca F and Leaver-Fay, Andrew and Jeliazkov, Jeliazko R and O’Meara, Matthew J and DiMaio, Frank P and Park, Hahnbeom and Shapovalov, Maxim V and Renfrew, P Douglas and Mulligan, Vikram K and Kappel, Kalli and others},
  journal={Journal of Chemical Theory and Computation},
  volume={13},
  number={6},
  pages={3031--3048},
  year={2017},  
}

@article{loffler2017rosetta,
  title={Rosetta: MSF: a modular framework for multi-state computational protein design},
  author={L{\"o}ffler, Patrick and Schmitz, Samuel and Hupfeld, Enrico and Sterner, Reinhard and Merkl, Rainer},
  journal={PLoS Computational Biology},
  volume={13},
  number={6},
  pages={e1005600},
  year={2017},
}

@inproceedings{jinglearning,
  title={Learning from protein structure with geometric vector perceptrons},
  author={Jing, Bowen and Eismann, Stephan and Suriana, Patricia and Townshend, Raphael John Lamarre and Dror, Ron},
  booktitle={International Conference on Learning Representations},
  year={2020},
}

@article{dauparas2022robust,
  title={Robust deep learning--based protein sequence design using ProteinMPNN},
  author={Dauparas, Justas and Anishchenko, Ivan and Bennett, Nathaniel and Bai, Hua and Ragotte, Robert J and Milles, Lukas F and Wicky, Basile IM and Courbet, Alexis and de Haas, Rob J and Bethel, Neville and others},
  journal={Science},
  volume={378},
  number={6615},
  pages={49--56},
  year={2022},
}

@inproceedings{gaopifold,
  title={PiFold: Toward effective and efficient protein inverse folding},
  author={Gao, Zhangyang and Tan, Cheng and Li, Stan Z},
  booktitle={International Conference on Learning Representations},
  year={2022}
}

@inproceedings{zheng2023structure,
  title={Structure-informed language models are protein designers},
  author={Zheng, Zaixiang and Deng, Yifan and Xue, Dongyu and Zhou, Yi and Ye, Fei and Gu, Quanquan},
  booktitle={International Conference on Machine Learning},  
  year={2023},  
}

@inproceedings{gaokw,
  title={KW-Design: Pushing the limit of protein design via knowledge refinement},
  author={Gao, Zhangyang and Tan, Cheng and Chen, Xingran and Zhang, Yijie and Xia, Jun and Li, Siyuan and Li, Stan Z},
  booktitle={International Conference on Learning Representations},
  year = {2023}
}

@inproceedings{yi2023graph,
  title={Graph denoising diffusion for inverse protein folding},
  author={Yi, Kai and Zhou, Bingxin and Shen, Yiqing and Li{\`o}, Pietro and Wang, Yuguang},
  booktitle={Advances in Neural Information Processing Systems},
  year={2023}
}

@article{bai2025mask,
  title={Mask-prior-guided denoising diffusion improves inverse protein folding},
  author={Bai, Peizhen and Miljkovi{\'c}, Filip and Liu, Xianyuan and De Maria, Leonardo and Croasdale-Wood, Rebecca and Rackham, Owen and Lu, Haiping},
  journal={Nature Machine Intelligence},
  pages={1--13},
  year={2025},  
}

@inproceedings{mahbub2025prism,
  title={PRISM: Enhancing protein inverse folding through fine-grained retrieval on structure-sequence multimodal representations},
  author={Mahbub, Sazan and Kundu, Souvik and Xing, Eric P},
  booktitle={International Conference on Learning Representations},
  year={2025}
}

@inproceedings{devlin2019bert,
  title={Bert: Pre-training of deep bidirectional transformers for language understanding},
  author={Devlin, Jacob and Chang, Ming-Wei and Lee, Kenton and Toutanova, Kristina},
  booktitle={Association for Computational Linguistics},
  year={2019}
}

@article{jumper2021highly,
  title={Highly accurate protein structure prediction with AlphaFold},
  author={Jumper, John and Evans, Richard and Pritzel, Alexander and Green, Tim and Figurnov, Michael and Ronneberger, Olaf and Tunyasuvunakool, Kathryn and Bates, Russ and {\v{Z}}{\'\i}dek, Augustin and Potapenko, Anna and others},
  journal={Nature},
  volume={596},
  number={7873},
  pages={583--589},
  year={2021},  
}

@article{akpinaroglu2023structure,
  title={Structure-conditioned masked language models for protein sequence design generalize beyond the native sequence space},
  author={Akpinaroglu, Deniz and Seki, Kosuke and Guo, Amy and Zhu, Eleanor and Kelly, Mark JS and Kortemme, Tanja},
  journal={bioRxiv},  
  year={2023},  
}

@inproceedings{satorras2021n,
  title={E (n) equivariant graph neural networks},
  author={Satorras, V{\i}ctor Garcia and Hoogeboom, Emiel and Welling, Max},
  booktitle={International Conference on Machine Learning},
  year={2021},  
}

@inproceedings{tan2023global,
  title={Global-context aware generative protein design},
  author={Tan, Cheng and Gao, Zhangyang and Xia, Jun and Hu, Bozhen and Li, Stan Z},
  booktitle ={International Conference on Acoustics, Speech and Signal Processing},  
  year={2023},
}

@inproceedings{ingraham2019generative,
  title={Generative models for graph-based protein design},
  author={Ingraham, John and Garg, Vikas and Barzilay, Regina and Jaakkola, Tommi},
  booktitle={Advances in Neural Information Processing Systems},  
  year={2019}
}

@article{gao2022alphadesign,
  title={Alphadesign: A graph protein design method and benchmark on alphafolddb},
  author={Gao, Zhangyang and Tan, Cheng and Li, Stan Z},
  journal={arXiv preprint arXiv:2202.01079},
  year={2022}
}

@inproceedings{hsu2022learning,
  title={Learning inverse folding from millions of predicted structures},
  author={Hsu, Chloe and Verkuil, Robert and Liu, Jason and Lin, Zeming and Hie, Brian and Sercu, Tom and Lerer, Adam and Rives, Alexander},
  booktitle={International Conference on Machine Learning},  
  year={2022},  
}

@article{orengo1997cath,
  title={CATH--a hierarchic classification of protein domain structures},
  author={Orengo, Christine A and Michie, Alex D and Jones, Susan and Jones, David T and Swindells, Mark B and Thornton, Janet M},
  journal={Structure},
  volume={5},
  number={8},
  pages={1093--1109},
  year={1997},
}

@article{li2014direct,
  title={Direct prediction of profiles of sequences compatible with a protein structure by neural networks with fragment-based local and energy-based nonlocal profiles},
  author={Li, Zhixiu and Yang, Yuedong and Faraggi, Eshel and Zhan, Jian and Zhou, Yaoqi},
  journal={Proteins: Structure, Function, and Bioinformatics},
  volume={82},
  number={10},
  pages={2565--2573},
  year={2014}
}

@article{zhou2023prorefiner,
  title={ProRefiner: an entropy-based refining strategy for inverse protein folding with global graph attention},
  author={Zhou, Xinyi and Chen, Guangyong and Ye, Junjie and Wang, Ercheng and Zhang, Jun and Mao, Cong and Li, Zhanwei and Hao, Jianye and Huang, Xingxu and Tang, Jin and others},
  journal={Nature Communications},
  volume={14},
  number={1},
  pages={7434},
  year={2023}
}

@article{berman2000protein,
  title={The protein data bank},
  author={Berman, Helen M and Westbrook, John and Feng, Zukang and Gilliland, Gary and Bhat, Talapady N and Weissig, Helge and Shindyalov, Ilya N and Bourne, Philip E},
  journal={Nucleic Acids Research},
  volume={28},
  number={1},
  pages={235--242},
  year={2000},  
}

@article{bairoch1997swiss,
  title={The SWISS-PROT protein sequence data bank and its supplement TrEMBL},
  author={Bairoch, Amos and Apweiler, Rolf},
  journal={Nucleic Acids Research},
  volume={25},
  number={1},
  pages={31--36},
  year={1997},
}

@article{henikoff1992amino,
  title={Amino acid substitution matrices from protein blocks.},
  author={Henikoff, Steven and Henikoff, Jorja G},
  journal={National Academy of Sciences},
  volume={89},
  number={22},
  pages={10915--10919},
  year={1992}
}

@article{passaro2025boltz2,
  author = {Passaro, Saro and Corso, Gabriele and Wohlwend, Jeremy and Reveiz, Mateo and Thaler, Stephan and Somnath, Vignesh Ram and Getz, Noah and Portnoi, Tally and Roy, Julien and Stark, Hannes and Kwabi-Addo, David and Beaini, Dominique and Jaakkola, Tommi and Barzilay, Regina},
  title = {Boltz-2: Towards accurate and efficient binding affinity prediction},
  year = {2025},
  journal = {bioRxiv}
}

@article{lin2022language,
  title={Language models of protein sequences at the scale of evolution enable accurate structure prediction},
  author={Lin, Zeming and Akin, Halil and Rao, Roshan and Hie, Brian and Zhu, Zhongkai and Lu, Wenting and dos Santos Costa, Allan and Fazel-Zarandi, Maryam and Sercu, Tom and Candido, Sal and others},
  journal={bioRxiv},
  year={2022}
}

@article{song2020denoising,
  title={Denoising diffusion implicit models},
  author={Song, Jiaming and Meng, Chenlin and Ermon, Stefano},
  journal={arXiv preprint arXiv:2010.02502},
  year={2020}
}

@inproceedings{austin2021structured,
  title={Structured denoising diffusion models in discrete state-spaces},
  author={Austin, Jacob and Johnson, Daniel D and Ho, Jonathan and Tarlow, Daniel and Van Den Berg, Rianne},
  booktitle={Advances in Neural Information Processing Systems},
  year={2021}
}

@article{litfin2025ultra,
  title={Ultra-fast and highly sensitive protein structure alignment with segment-level representations and block-sparse optimization},
  author={Litfin, Thomas and Zhou, Yaoqi and von Itzstein, Mark},
  journal={bioRxiv},  
  year={2025},  
}

@inproceedings{huang2024interaction,
  title={Interaction-based retrieval-augmented diffusion models for protein-specific 3d molecule generation},
  author={Huang, Zhilin and Yang, Ling and Zhou, Xiangxin and Qin, Chujun and Yu, Yijie and Zheng, Xiawu and Zhou, Zikun and Zhang, Wentao and Wang, Yu and Yang, Wenming},
  booktitle={International Conference on Machine Learning},
  year={2024}
}

@article{hayes2025simulating,
  title={Simulating 500 million years of evolution with a language model},
  author={Hayes, Thomas and Rao, Roshan and Akin, Halil and Sofroniew, Nicholas J and Oktay, Deniz and Lin, Zeming and Verkuil, Robert and Tran, Vincent Q and Deaton, Jonathan and Wiggert, Marius and others},
  journal={Science},
  volume={387},
  number={6736},
  pages={850--858},
  year={2025},
}

@article{koehler2023sequence,
  title={Sequence-structure-function relationships in the microbial protein universe},
  author={Koehler Leman, Julia and Szczerbiak, Pawel and Renfrew, P Douglas and Gligorijevic, Vladimir and Berenberg, Daniel and Vatanen, Tommi and Taylor, Bryn C and Chandler, Chris and Janssen, Stefan and Pataki, Andras and others},
  journal={Nature Communications},
  volume={14},
  number={1},
  pages={2351},
  year={2023},
}

@article{ganea2021independent,
  title={Independent se (3)-equivariant models for end-to-end rigid protein docking},
  author={Ganea, Octavian-Eugen and Huang, Xinyuan and Bunne, Charlotte and Bian, Yatao and Barzilay, Regina and Jaakkola, Tommi and Krause, Andreas},
  journal={arXiv preprint arXiv:2111.07786},
  year={2021}
}

@article{wang2024diffusion,
  title={Diffusion language models are versatile protein learners},
  author={Wang, Xinyou and Zheng, Zaixiang and Ye, Fei and Xue, Dongyu and Huang, Shujian and Gu, Quanquan},
  journal={arXiv preprint arXiv:2402.18567},
  year={2024}
}

@article{wang2018computational,
  title={Computational protein design with deep learning neural networks},
  author={Wang, Jingxue and Cao, Huali and Zhang, John ZH and Qi, Yifei},
  journal={Scientific Reports},
  volume={8},
  number={1},
  pages={1--9},
  year={2018},
}

@article{qi2020densecpd,
  title={DenseCPD: improving the accuracy of neural-network-based computational protein sequence design with DenseNet},
  author={Qi, Yifei and Zhang, John ZH},
  journal={Journal of Chemical Information and Modeling},
  volume={60},
  number={3},
  pages={1245--1252},
  year={2020},  
}

@article{qiu2024instructplm,
  title={Instructplm: Aligning protein language models to follow protein structure instructions},
  author={Qiu, Jiezhong and Xu, Junde and Hu, Jie and Cao, Hanqun and Hou, Liya and Gao, Zijun and Zhou, Xinyi and Li, Anni and Li, Xiujuan and Cui, Bin and others},
  journal={bioRxiv},
  year={2024},
}

@article{varadi2022alphafold,
  title={AlphaFold Protein Structure Database: massively expanding the structural coverage of protein-sequence space with high-accuracy models},
  author={Varadi, Mihaly and Anyango, Stephen and Deshpande, Mandar and Nair, Sreenath and Natassia, Cindy and Yordanova, Galabina and Yuan, David and Stroe, Oana and Wood, Gemma and Laydon, Agata and others},
  journal={Nucleic Acids Research},
  volume={50},
  number={D1},
  pages={D439--D444},
  year={2022},
}

@inproceedings{nichol2021improved,
  title={Improved denoising diffusion probabilistic models},
  author={Nichol, Alexander Quinn and Dhariwal, Prafulla},
  booktitle={International Conference on Machine Learning},
  year={2021},
}
\bibliographystyle{iclr2026_conference}

\newpage 
\appendix

\end{document}